\documentclass[aps,pra,preprint,superscriptaddress,nofootinbib]{revtex4}

\usepackage{amssymb,amsmath,graphicx}

\begin{document}

\title{Nonlocal Quantum Effects in Cosmology}

\author{Yurii V. Dumin}

\email[E-mail: ]{dumin@yahoo.com, dumin@sai.msu.ru}

\affiliation{
Sternberg Astronomical Institute (GAISh) of \\
Lomonosov Moscow State University, \\
Universitetski prosp.\ 13, 119991, Moscow, Russia}
\affiliation{
Space Research Institute (IKI) of \\
Russian Academy of Sciences, \\
Profsoyuznaya str.\ 84/32, 117997, Moscow, Russia}

\date{January 7, 2014; Revised: April 7, 2014}

\begin{abstract}
Since it is commonly believed that the observed large-scale
structure of the Universe is an imprint of quantum fluctuations
existing at the very early stage of its evolution, it is reasonable
to pose the question:
Do the effects of quantum nonlocality, which are well established
now by the laboratory studies, manifest themselves also in the early
Universe?
We try to answer this question by utilizing the results of a few
experiments, namely, with the superconducting multi-Josephson-junction
loops and the ultracold gases in periodic potentials.
Employing a close analogy between the above-mentioned setups and
the simplest one-dimensional Friedmann--Robertson--Walker
cosmological model, we show that the specific nonlocal correlations
revealed in the laboratory studies might be of considerable importance
also in treating the strongly-nonequilibrium phase transitions
of Higgs fields in the early Universe.
Particularly, they should substantially reduce the number of
topological defects (\textit{e.g.}, domain walls) expected due to
independent establishment of the new phases in the remote spatial
regions. This gives us a hint for resolving a long-standing problem of
the excessive concentration of topological defects, inconsistent with
observational constraints.
The same effect may be also relevant to the recent problem of
the anomalous behavior of cosmic microwave background fluctuations
at large angular scales.
\end{abstract}

\maketitle

\section*{\large 1. Introduction}

\begin{figure}[b]
\includegraphics[width=11.5cm]{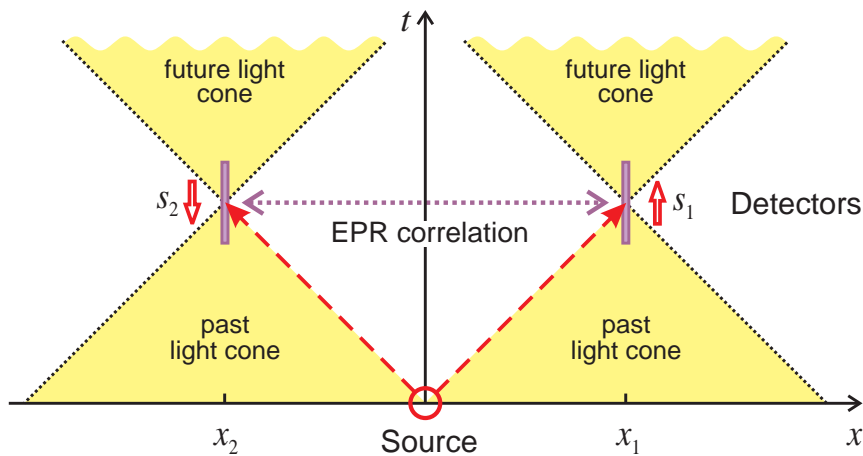}
\caption{\label{fig:EPR_lab}
Sketch of the typical laboratory ERP experiment.}
\end{figure}

The concept of quantum nonlocality originates actually from
the paper by Einstein, Podolsky, and Rosen (EPR)~\cite{Einstein35},
who posed the problem of correlation between the measurements of
two physical objects located in the causally-disconnected
regions of space, \textit{i.e.}, beyond the light cones of each other.
In the modern and most frequently used in the experiments form,
this phenomenon can be illustrated in Figure~\ref{fig:EPR_lab}.
Here, an original particle of zero spin decays
at the instant $ t = 0 $ into two particles with equal but
oppositely-directed spins $ s_1 $ and $ s_2 $, which
subsequently move from each other in the opposite directions.
Next, if measurements of the spins of both particles are
performed in the remote spatial points $ x_1 $ and $ x_2 $
at the same instant of time, their values turn out to be
perfectly correlated ($ s_1 = - s_2 $) just because of
the law of spin conservation.

At first sight, such correlation is quite surprising, because
the measurements are performed in the spots of space--time
lying beyond the mutual light cones (\textit{i.e.}, in
the causally disconnected regions). However, the existence of
EPR correlations is well confirmed now by a lot of laboratory
experiments. In fact, these correlations look much less
surprising if we keep in mind the fact that both light cones
include the same common source in the past.
It might be reasonable to emphasize also that, despite of
a ``superluminal'' character of the EPR correlations,
they cannot be employed for a faster-than-light communication,
because the outcomes of correlated measurements of $ s_1 $
and $ s_2 $ in the points $ x_1 $ and $ x_2 $ are random.

Turning attention to cosmology, we should first of all
mention that it is widely believed now that the observed
large-scale structure of the Universe is an imprint of
quantum fluctuations existing at the very early stages of
cosmological evolution.
Therefore, it becomes interesting to pose the question:
Can the effects of quantum nonlocality, similar to
EPR correlations, manifest themselves also in cosmology?
In particular, it should be kept in mind that Higgs field,
filling the entire Universe and giving masses to the elementary
particles, represents actually a kind of Bose--Einstein
condensate (BEC), \textit{i.e.}, a macroscopic quantum state,
in which the specific quantum correlations may naturally occur.

The most important feature in temporal dynamics of the Higgs field
is phase transition caused by the evolving temperature
of the Universe, which can finally result in the formation of
the nontrivial states of the physical vacuum~\cite{Linde79}.
In fact, the problem of complex vacuum was recognized long
time ago, just after appearance of the idea of spontaneous
symmetry breaking in the quantum field theory.
Particularly, at the Conference on the occasion of
the 400th anniversary of Galileo Galilei's birth, held in Pisa
in 1964, Bogoliubov emphasized that ``it is hard to admit,
for example, that the `phases' are the same everywhere in
the space. So it appears necessary to consider such things as
`domain structure' of the vacuum''~\cite{Bogoliubov66}.
In the next decade, the problem of formation of the domain walls
in the course of cosmological evolution was considered in much
detail in the work by Zeldovich, Kobzarev, and
Okun~\cite{Zeldovich74} and later in papers by many other
authors. (A quite comprehensive overview of the domain wall
dynamics was given, for example, in paper~\cite{Gelmini89}.)

\begin{figure}
\includegraphics[width=13cm]{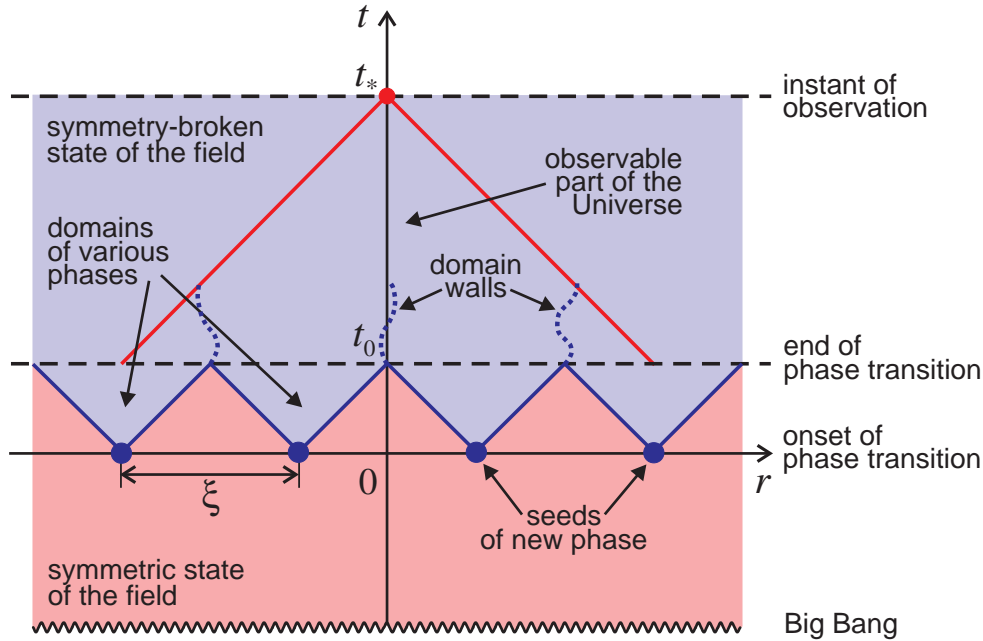}
\caption{\label{fig:Cosm_evol}
Sketch of development of the phase transition
in the expanding Universe.}
\end{figure}

A commonly-accepted scenario of formation of the domain walls
by the phase transition in the early Universe is illustrated in
Figure~\ref{fig:Cosm_evol}.
A uniform initial (``symmetric'') state of the Higgs field,
existing soon after the Big Bang, cools down due to expansion
of the Universe; so that the symmetric phase of the field
becomes energetically unfavorable, and the seeds of a new
(``symmetry-broken'') phase emerge in some spatial points.
(For the sake of simplicity, we shall assume that
these points are equally separated along the coordinate~$ r $
and originate at the same instant of time $ t = 0 $.)
Since the seeds of the low-temperature phase
emerge independently in the remote spatial regions,
their states of the degenerate vacuum, in general, will be
different from each other.

Then, the domains of the new phase quickly grow in the course
of time and at the instant $ t = t_0 $ collide with each other.
However, if the values of the symmetry-broken phase in
two neighboring domains were different,
they cannot merge smoothly. Instead, a stable topological
defect---domain wall (or ``kink'')---should be formed at
their boundary. So, if the initial separation between
the seeds of the new phase was~$ \xi $, then the resulting
concentration of the domain walls can be roughly estimated as
$ n \approx 1/{\xi} $. This is the so-called
Kibble--Zurek (KZ) mechanism for the formation of
topological defects after the strongly-nonequilibrium phase
transformations~\cite{Kibble76,Zurek85}.

Strictly speaking, the scenario outlined in Figure~\ref{fig:Cosm_evol}
refers to the simplest case of Higgs field, possessing
$ {\mathbb{Z}}_2 $ symmetry group (\textit{i.e.}, admitting
a discrete symmetry breaking). However, the same basic idea
is applicable also to more realistic Higgs fields with
the continuous symmetries, whose breaking can lead to
the formation of more complex defects of the vacuum,
such as the monopoles and cosmic strings (or vortices).
In general, KZ mechanism gives the following estimate for
concentration of the topological defects:
\begin{equation}
n \approx 1 / \, {\xi}_{\rm eff}^{\,d} \, ,
\label{eq:KZ_concentr}
\end{equation}
where $ {\xi}_{\rm eff} $ is the effective correlation length,
\textit{i.e.}, a typical distance between the seeds of
the new phase;
and $ d = $~3, 2, and 1 for the monopoles, cosmic strings, and
domain walls, respectively.
(Their concentrations refer evidently to unit volume, area,
and length.)

An accurate calculation of the effective correlation length,
in general, is a difficult task.
However, its upper bound can be obtained from a simple
causality argument:
$
{\xi}_{\rm eff} \leq c \, t_{\rm pt}
$,
where $ c $~is the speed of light, and
$ t_{\rm pt} $~is the characteristic time from the Big Bang to
the instant of phase transition.
Next, the time from the Big Bang in Friedmann--Robertson--Walker (FRW)
cosmology is estimated as $ {\sim}1/H $, where $ H $~is the value of
Hubble parameter at the respective instant.
Consequently,
\begin{equation}
{\xi}_{\rm eff} \, \lesssim \, c \, / H_{\rm pt} \, .
\label{eq:xi_causal_est}
\end{equation}
At last, substituting~(\ref{eq:xi_causal_est})
into~(\ref{eq:KZ_concentr}), we get:
\begin{equation}
n \, \gtrsim \, \left( H_{\rm pt} / c \right){}^d \, ,
\label{eq:Conc_lower_bound}
\end{equation}
where $ H_{\rm pt} $ is the value of Hubble parameter at
the instant of phase transition.%
\footnote{
To avoid misunderstanding, let us mention that most of
laboratory experiments aimed at verification of KZ mechanism
measured a scaling relationship between the size of uniform
domains and the quench rate, rather than the absolute number
of the defects. However, in cosmological applications it is
more appropriate to discuss just the absolute concentration
of the defects.}

Unfortunately, the lower theoretical bound~(\ref{eq:Conc_lower_bound})
is inconsistent with the upper bounds following from observations
(\textit{e.g.}, review~\cite{Klapdor97}).
One possible way to mitigate this disagreement is to modify
Lagrangian of the field theory under consideration,
typically, by introduction of the ``biased'' vacuum (thereby,
\textit{a priori} removing the degeneracy)~\cite{Zeldovich74,Gelmini89}.
Yet another conceivable approach, which was not exploited
before, is to take into account the nonlocal quantum
correlations, which might manifest themselves in
the macroscopic BEC.

However, exploiting the idea of macroscopic quantum
correlations, one should bear in mind the following
two subtle points:
\begin{itemize}

\item Firstly, the most of laboratory studies of EPR
correlations, performed since the 1970's, dealt with
the microscopic quantum objects (\textit{e.g.}, atoms
and photons). There were only a very few number of experiments
on the nonlocal correlations in macroscopic systems;
and they will be discussed in detail in the next section.

\item Secondly, the correlations studied in microscopic objects
were always associated with the exact conservation laws
(most typically, the conservation of the total spin).
In contract to these cases, the macroscopic BECs do not
usually obey the suitable conservation laws.
Nevertheless, it might be expected that correlations in
the macroscopic systems could be caused just by the energetic
criteria: if correlated state of a large system possesses
less energy than its uncorrelated state, then it should emerge
with a greater probability.
As will be shown in the next section, a number of recent
laboratory experiments support this idea.

\end{itemize}

\section*{\large 2. Review of the Laboratory Experiments}

As far as we know, there are by now two groups of experiments
confirming the phenomenon of nonlocal correlations in
macroscopic BECs.
The first of them are the experiments with superconducting
multi-Josephson-junction loops (MJJL), which were started in
the very beginning of 2000's~\cite{Carmi00}; and the second are
the experiments with ultracold gases in the periodic optical potentials,
which began a few years later~\cite{Hadzibabic04,Hadzibabic06}.

\subsection*{\large \it 2.1. MJJL Experiment}

\begin{figure}
\includegraphics[width=9cm]{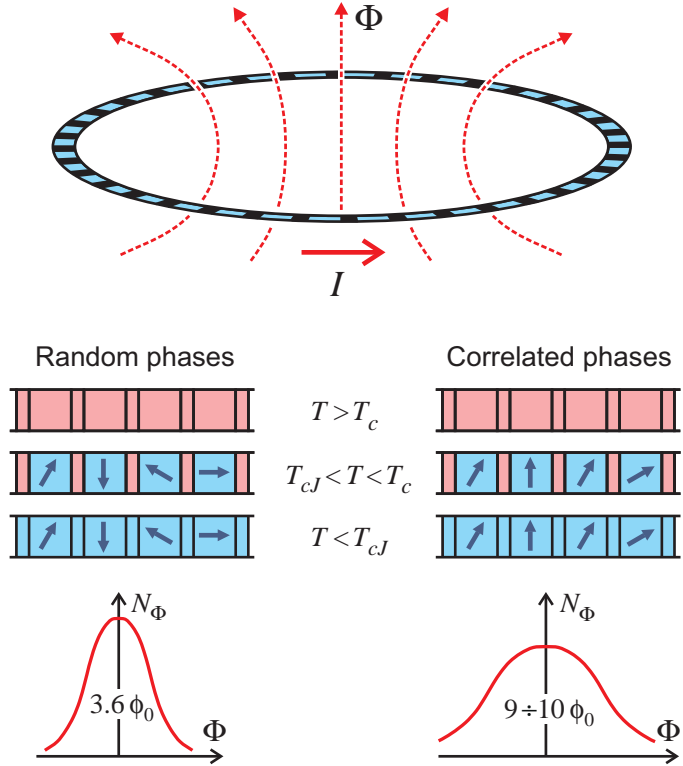}
\caption{\label{fig:MJJL_exp}
Basic design and principal results of the MJJL experiment.}
\end{figure}

A general scheme of the original MJJL experiment~\cite{Carmi00}
is shown in Figure~\ref{fig:MJJL_exp}:
A thin quasi-one-dimensional loop was fabricated from
the YBa${}_2$Cu${}_3$O${}_{7\text{-}\delta}$ superconductor
and contained 214~segments separated from each other by
the Josephson junctions (which are the microscopic
domains of the same substance but with a higher critical
temperature due to defects in the crystalline lattice).%
\footnote{
The actual loop used in the experiment was not perfectly
circular, as in figure, but represented a winding strip
engraved in the superconductor film.}
The experimental procedure consisted of the numerous cycles
of very quick cooling (approximately from 100~K to 77~K)
and subsequent heating of the loop and measurement of
the resulting electromagnetic response.

In the phase of cooling, when temperature~$ T $ drops approximately
to $T_c = 90$~K, the segments separated by the junctions become
superconducting. However, the junctions themselves remain
normal and, therefore, the superconducting segments
are effectively separated from each other.
Consequently, a random phase of the superconducting order parameter
should be established in each of them.
(They are schematically illustrated in Figure~\ref{fig:MJJL_exp}
by the randomly oriented arrows.)
The total phase variation of the order parameter along the loop,
in general, should be nonzero.

At subsequent cooling down to the temperature $T_{cJ}$, which is
$5{\div}7$~K below $T_c$, the Josephson junctions become also
superconducting. Consequently, due to the above-mentioned phase
variation, a superconducting current~$ I $ will develop along the loop.
As a result, the loop will be penetrated by the magnetic flux~$ \Phi $,
which is just the measurable quantity.
It should be mentioned that this setup is quite close to the original
idea by Zurek~\cite{Zurek85}, who proposed to observe a spontaneous
rotation produced by a rapid phase transition to the superfluid phase
in a thin annular tube.
Unfortunately, such an experiment was never implemented in practice,
since it is hardly possible to observe a mechanical rotation of
the macroscopic body with angular momentum of just a few quanta~$ \hbar $.
On the other hand, the magnetic flux measurements by modern apparatus
can easily detect the individual quanta of the magnetic flux.
Therefore, an electromagnetic analog of the original Zurek's proposal
became feasible.

In summary, because of the phase jumps between the isolated
segments formed at the stage when $ T_{cJ} < T < T_c $, the final magnetic
flux~$ \Phi $ through the loop turns out to be nonzero and varies randomly
from one heating--cooling cycle to another.
The histogram $ N_{\Phi}(\Phi) $ derived from a large number of cycles
is well described by the normal (Gaussian) law with zero average value
and standard deviation $7.4 \, {\phi}_0$ (where $ {\phi}_0 $~is
the magnetic flux quantum, and $ N_{\Phi} $~is the number of cases with
the total magnetic flux~$ \Phi $).
This standard deviation is just the typical value of the flux
spontaneously generated in one cycle.

In fact, the above-written value is unreasonably large: if the phase
jumps between the segments were absolutely independent of each other,
then the expected width of the distribution would be
only~$3.6 \, {\phi}_0$.
However, the excessive value was satisfactorily explained
by the authors of the experiment under assumption that phases of
the superconducting order parameter in the isolated
(\textit{i.e.}, ``causally-disconnected'') segments were correlated to
each other, so that probability $ P({\delta}_i) $ of the phase
jump~$ {\delta}_i $ in the $i$'th junction was given by the Gibbs law:
\begin{equation}
P({\delta}_i) \propto \exp [-{E_J}({\delta}_i) / k_B T] \, ,
\label{eq:MJJL_corr}
\end{equation}
where
$ E_J $~is the energy concentrated in the Josephson junction,
$ T $~is the temperature, and
$ k_B $~is Boltzmann constant.

So, the main conclusion following from the above experiment is that
the energy concentrated in the defects should be taken into account
in the calculation of the probability of realization of various field
configurations, even if the phase transformation develops independently
in the remote parts of the system.

\subsection*{\large \it 2.2. Experiments with Ultracold Gases
in Optical Lattices}

The MJJL experiment, discussed in the previous section, gave the first hint
to the importance of using the Gibbs law even for the systems composed
of the apparently isolated parts.
Unfortunately, this experiment did not enable us to check the particular
functional dependence~(\ref{eq:MJJL_corr}).
(To do so, it would be necessary to perform the same experiment with
different types of superconductors, which has not been fulfilled by now.)

Nevertheless, a few yeas later it became possible to solve this tack
by using the BECs of ultracold gases in periodic potentials (or
the so-called optical lattices), formed by the intersecting laser beams.
These systems represent a close analog of the multiple Josephson junctions
and, as distinct from the solid-state setups, their parameters can be
easily varied. A diagnostics of the phase jumps in such installations is
performed by a removal of the external potential, thereby enabling
the pieces of BEC to expand and interfere with each other.

For example, an array of 30~BECs of the ultracold gas in a regular
one-dimensional lattice was created in the experiment~\cite{Hadzibabic04}.
Next, it was demonstrated that such condensates can well interfere
with each other even if they were produced independently,
\textit{i.e.}, ``have never seen one another''.

\begin{figure}
\includegraphics[width=9cm]{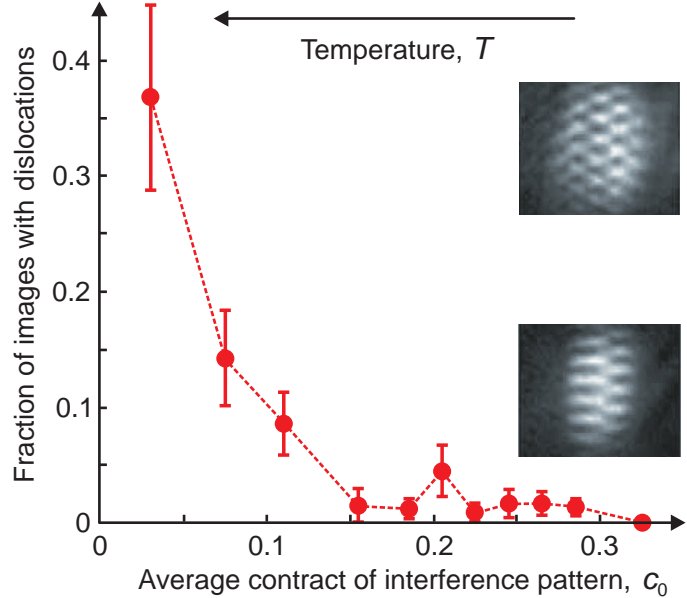}
\caption{\label{fig:Defects_in_Gas}
Fraction of the interference patterns showing at least one dislocation
as function of the interference contrast~$ c_0 $.
Inset, examples of images with the increasing number of dislocations
(from bottom to top).
Adapted by permission from Macmillan Publishers Ltd:
\emph{Nature}, vol.~441, no.~7097, pp.~1118--1121, \copyright 2006.}
\end{figure}

A further experiment of the same group~\cite{Hadzibabic06} was devoted to
a detailed study of the phase defects. In particular, an efficiency of
the defect formation was measured as function of temperature.
(In fact, the determination of temperature is not an easy task in the
experiments with ultracold gases. So, the primary independent parameter
was taken to be the average contract of the interference pattern~$ c_0 $,
which is, roughly speaking, inversely proportional to the temperature~$ T $.)
As a result, it was found that the number of defects (dislocations)
formed in the BEC of ultracold gas increases with temperature by
qualitatively the same law as~(\ref{eq:MJJL_corr});
see Figure~\ref{fig:Defects_in_Gas} (a sharp outlier at
$ c_0 \approx 0.2 $ is most probably an experimental inaccuracy).

Therefore, both the MJJL and ultracold-atom experiments suggest that
the probability of realization of various configurations of the BEC
order parameter should be calculated taking into account
\textit{the total energy of the system} even if separate parts of
this system do not interact with each other during a particular
physical process (\textit{e.g.}, the phase transformation).
Of course, these parts of the system must be causally connected
during its previous evolution. This is always satisfied in
the laboratory experiments but requires a special consideration in
the cosmological context.

In some sense, the above phenomenon can be interpreted as analog of
EPR correlation for the system that does not posses an exact
conservation law. In such a case, just the energetic criteria
should come into play (see also discussion in the end of Section~1).

\section*{\large 3. Cosmological Implications}

\begin{figure}
\includegraphics[width=7cm]{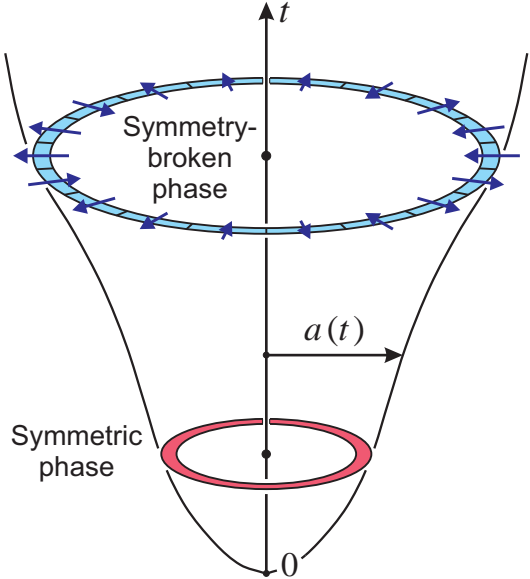}
\caption{\label{fig:1D_FRW}
Sketch of the phase transition in 1D FRW cosmological model
(the physical distance is measured along the circles).}
\end{figure}

There is evidently a close similarity between the symmetry-breaking phase
transitions in the multi-Josephson-junction loop, depicted in
Figure~\ref{fig:MJJL_exp}, and in the simplest one-dimensional (1D)
Friedmann--Robertson--Walker (FRW) cosmological model, schematically
illustrated in Figure~\ref{fig:1D_FRW}.
For the sake of definiteness, we shall consider only the simplest type
of defects, namely, the domain walls or kinks.

To make the quantitative estimates, let us consider the space--time metric
\begin{equation}
ds^2 = \: dt^2 - \, a^2(t) \: dx^2 \, ,
\label{eq:metric}
\end{equation}
where $ t $~is the time, $ x $~is the spatial coordinate,
and $a(t)$~is the scale factor of FRW model.
(From here on, we shall assume that $ c \! \equiv \! 1 $.)
Let this space--time be filled with the real scalar field~$ \varphi $,
simulating Higgs field in the theory of elementary particles.
Its Lagrangian
\begin{equation}
{\cal L} \, (x, t) \, = \,
\frac{1}{2} \, \big[ {\left( {\partial}_t \varphi \right)}^2 \! - \,
  {\left( {\partial}_x \varphi \right)}^2 \big]
  \, - \:
\frac{\lambda}{4} \,
  {\big[ \, {\varphi}^2 \! - \left( {\mu}^2 / \lambda \right) \big]}^2
\label{eq:Lagrangian}
\end{equation}
possesses $ \mathbb{Z}_2 $ symmetry group, which should be broken by
the phase transition.

As is known, the stable low-temperature vacuum states
of the field~(\ref{eq:Lagrangian}) are
\begin{equation}
{\varphi}_0 = \, \pm \, \mu \, / \sqrt{\lambda} \: ,
\end{equation}
and a transition region between them (domain wall) is described as
\begin{equation}
\varphi \, (x) = \, {\varphi}_0 \tanh \! \big[ ({\mu} / {\sqrt{2}}\,) \,
  ( x - x_0 ) \big] \, .
\end{equation}

Such a domain wall contains the energy
\begin{equation}
E = \, \frac{2 \, \sqrt{2}}{3} \, \frac{{\mu}^3}{\lambda} \; .
\label{eq:d-w_energy}
\end{equation}
We shall assume that thickness of the wall, $ {\sim}{1 / \mu} $,
is small in comparison with a characteristic distance between them;
\textit{i.e.}, the domain walls can be treated as point-like objects.

Next, it is convenient to introduce the conformal time
$ \eta = \! \int dt / a(t) \, $.
As a result, the space--time metric~(\ref{eq:metric}) will take
the conformally flat form~\cite{Misner69}:
\begin{equation}
ds^2 = \, a^2(t) \, [ \, d{\eta}^2 \! - dx^2 \, ] \, ;
\end{equation}
so that the light rays ($ ds^2 \! = 0 $) will represent the straight lines
inclined at $ \pm \pi / 4 \, $:
\begin{equation}
x = \pm \, \eta + {\rm const} \, .
\end{equation}

\begin{figure}
\includegraphics[width=9.5cm]{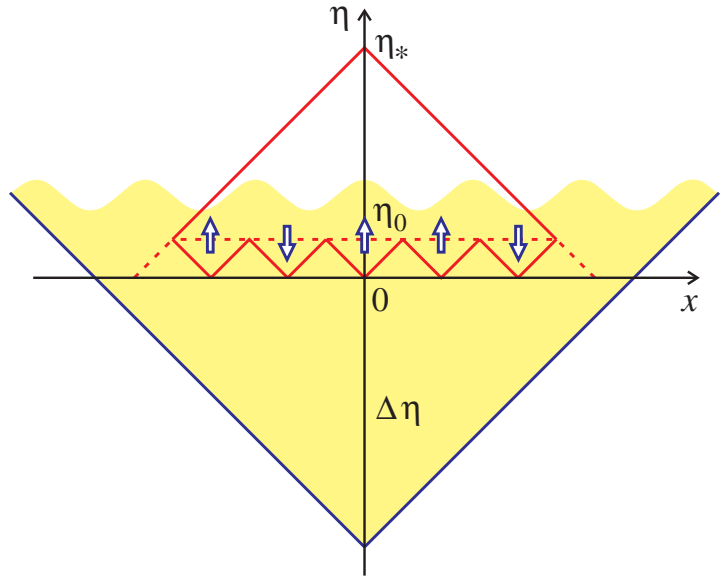}
\caption{\label{fig:conf_diag}
Conformal diagram of 1D FRW cosmological model.}
\end{figure}

The entire structure of the space--time can be conveniently described by
the conformal diagram in Figure~\ref{fig:conf_diag}.
Let $ \eta \! = \! 0 $ and $ \eta \! = \! {\eta}_0 $ be the beginning and
end of the phase transition, respectively, and $ \eta \! = \! {\eta}_* $ be
the instant of observation.%
\footnote{
The instants $ \eta \! = 0 $, $ {\eta}_0 $, and $ {\eta}_* $ of
the conformal time correspond to the instants
$ t \! = 0 $, $ t_0 $, and $ t_* $ of the physical time
in Figure~\ref{fig:Cosm_evol}.}
Since it is commonly assumed that bubbles of the new phase grow at the rate
close to the speed of light, their boundaries can be well depicted by
the light rays.
Then, as follows from a simple geometric consideration,
\begin{equation}
N \, = \, ( {\eta}_* \! - {\eta}_0 ) / {\eta}_0 \,
  \approx \, {\eta}_* / {\eta}_0
\quad (\text{at large } N)
\label{eq:num_subreg}
\end{equation}
is the number of spatial subregions in the observable Universe
causally-disconnected \textit{during} the phase transition.
Their final vacuum states can be conveniently denoted by
the arrows, like spins.

Let us calculate a probability of the phase transition without
formation of the domain walls in the observable space--time
(the past light cone)~$ P_N^{\,0} $, where subscript~$ N $ implies
the number of subregions; and superscript~0, the absence of
domain walls.
A trivial estimate can be obtained by taking a ratio of the number
of field configurations without domain walls~(which equals~2) to
their total number~($ 2^N $):
\begin{equation}
P_N^{\,0} = \, 2 \, / \, 2^N \, = \, 1 \, / \, 2^{N-1} .
\label{eq:triv_probabil}
\end{equation}
This quantity evidently tends to zero very quickly at large~$ N $.
In other words, the observable part of the Universe, represented by
the large upper triangle in Figure~\ref{fig:conf_diag},
will inevitably contain some number of the domain walls.
Unfortunately, as was recognized long time ago~\cite{Zeldovich74,Gelmini89},
a presence of the domain walls is incompatible with astronomical
observations.

A possible resolution of this paradox can be based just on taking into
consideration the nonlocal Gibbs-like correlations~(\ref{eq:MJJL_corr}).
First of all, we must ensure that such correlations can really develop,
\textit{i.e.}, the subdomains of the new phase were causally connected
in the past.
As is seen in Figure~\ref{fig:conf_diag}, this is really possible if
a sufficiently long interval of the conformal time
\begin{equation}
\Delta \eta \, \ge \, {\eta}_*
\label{eq:entangl_state}
\end{equation}
preceded the phase transition. Then, the lower shaded triangle
will cover at the instant~$ \eta \! = \! 0 $ the upper triangle,
representing the observable part of the Universe.

The inequality~(\ref{eq:entangl_state}) can be satisfied, particularly,
in the case of sufficiently long inflationary (de~Sitter) stage.
Really, if
$ \, a(t) \propto \exp (Ht) $,
where~$ H $~is Hubble constant, then
\begin{equation}
\eta \, \propto \, - \, \frac{1}{H} \, e^{-Ht} + {\rm const} \,
  \to \, - \infty \quad \text{at} \quad t \, \to \, - \infty \, ;
\end{equation}
so that the above-mentioned interval~$ \Delta \eta $ can be sufficiently
large.
Let us remind that, from the viewpoint of elementary-particle physics,
the de~Sitter stage can be naturally realized in the overcooled state of
the Higgs field immediately before its first-order phase transition;
and just this idea was the starting point of the first inflationary
models~\cite{Linde84}.

Next, if the condition~(\ref{eq:entangl_state}) is satisfied,
then it is reasonable to assume that the above-mentioned
correlations~(\ref{eq:MJJL_corr}) should take place between the all
$ N $ subdomains drawn in the conformal diagram, Figure~\ref{fig:conf_diag}.
(It is interesting to mention that in our old work~\cite{Dumin00},
performed before the MJJL experiment, the same Gibbs-like correlations
were introduced on the basis of some metaphysical speculations.)
In such a case, the probability $ P_N^{\,0} $ should be calculated taking
into account Gibbs factors for the field configurations involving
the domain walls:
\begin{equation}
P_N^{\,0} \, = \, 2 \, / \, Z \: ,
\end{equation}
where
\begin{equation}
Z \, = \: \sum_{i=1}^N \;
  \sum_{s_i = \pm 1}
  \exp \bigg\lbrace \!
    - \frac{E}{T} \, \sum_{j=1}^N \,
    \frac{1}{2} \, (1 - \, s_{j} \, s_{j+1})
  \bigg\rbrace \, .
\label{eq:stat_sum}
\end{equation}
Here, $ s_j $~is the spin-like variable denoting a sign of the vacuum state
in the $j$'th subdomain,
$ E $~is the domain wall energy, given by~(\ref{eq:d-w_energy}),
and $ T $~is the characteristic temperature of the phase transition.
(From here on, the temperature will be expressed in energetic units;
so the Boltzmann constant~$ k_B $ will be omitted.)

From a formal point of view, statistical sum~(\ref{eq:stat_sum}) is
very similar to the sum for Ising model, well studied in the physics
of condensed matter, \textit{e.g.}~\cite{Isihara71}.
Using exactly the same mathematical approach,
we get the final result:%
\footnote{Attention should be paid to the appropriate choice of zero
energy, which is different from the one commonly used in
the condensed-matter physics.}
\begin{equation}
P_N^{\,0} \, = \, \frac{2}{
  {\left[ 1 + e^{- E / T} \right]}^N +
  {\left[ 1 - e^{- E / T} \right]}^N } \; .
\label{eq:Gibbs_probabil}
\end{equation}
(Yet another method for calculation of the same quantity, which is more
straightforward and pictorial but less informative, can be found
in~\cite{Dumin00}; the approach outlined here was employed for the first
time in our paper~\cite{Dumin03}.)

\begin{figure}
\includegraphics[width=8cm]{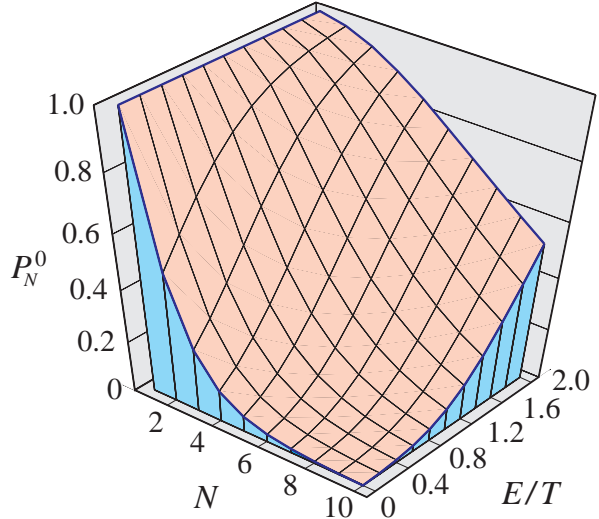}
\caption{\label{fig:prob_distr}
The probability of phase transition without formation of
the domain walls~$ P_N^{\,0} $ as function of the number of
disconnected subregions~$ N $ and the ratio of the domain wall
energy to the phase transition temperature~$ E/T $.}
\end{figure}

The quantity~$ P_N^{\,0} $ as function of~$ N $ and~$ E/T $ is plotted
in Figure~\ref{fig:prob_distr}.
It is seen that $ P_N^{\,0} $ drops very sharply with increase in~$ N $
at small values of $ E/T $ (when the effect of Gibbs-like correlations
is insignificant), but it becomes a gently decreasing function of~$ N $
when the parameter~$ E/T $ is sufficiently large.
Some other plots illustrating suppression of concentration of the domain
walls by the nonlocal correlations can be found in paper~\cite{Dumin09},
devoted to the phase transformations in superfluids and superconductors.
Therefore, just the large energy concentrated in the domain walls
turns out to be the factor substantially reducing the probability of
their creation.

Next, as can be easily derived from~(\ref{eq:Gibbs_probabil}),
the probability of absence of the domain walls in the observable Universe
becomes on the order of unity, \textit{e.g.}\ 1/2, if
$\, E/T \gtrsim \ln N $.
Taking into account~(\ref{eq:d-w_energy}) and~(\ref{eq:num_subreg}),
this inequality can be rewritten as
\begin{equation}
\frac{{\mu}^3}{\lambda \, T} \: \gtrsim \:
  \ln \frac{{\eta}_*}{{\eta}_0} \: .
\label{eq:d-w_absence}
\end{equation}
Because of the very weak logarithmic dependence in the right-hand side,
such a condition could be reasonably satisfied for a certain class of
field theories.

Moreover, the situation becomes even more favorable in the case of
two- or three-dimensional space. The point is that a well-known
property of the 2D and 3D Ising models is a tendency for aggregation
of the domains with the same value of the order parameter when
the temperature drops below some critical value
$ T_c \sim E $~\cite{Rumer80}.
In the condensed-matter applications, this corresponds, for example,
to the spontaneous magnetization of a solid body.
Regarding the cosmological context, one can expect that probability of
formation of the domain walls will be reduced dramatically at
the sufficiently large values of $ E/T $; some illustrations of this
phenomenon can be found in~\cite{Dumin09}.
(To avoid misunderstanding, let us emphasize that the above-mentioned
Ising models are only the auxiliary mathematical constructions,
describing a final distribution of the domain walls after the phase
transformation. So, the formal phase transitions in the 2D and 3D Ising
models should not be associated with the physical phase transition in
the original $ {\varphi}^4 $-field model~(\ref{eq:Lagrangian});
for more details, see Table~2 in~\cite{Dumin09}.)

\section*{\large 4. Discussion and Conclusions}

As was discussed in the present article, a few laboratory experiments
suggest a presence of the nonlocal Gibbs-like correlations between
the phases of BECs after the rapid phase transformations.
Therefore, it can be reasonably conjectured that the same correlations
should occur in the BEC of Higgs field, which is formed in the course of
evolution of the Universe.
As follows from our quantitative estimates for the simplest case of
1D FRW cosmological model, such correlations can show a way to resolve
the well-known problem of the excessive concentration of the domain walls
resulting from phase transitions in the early Universe.
In fact, this problem was recognized in the mid 1970s~\cite{Zeldovich74};
and since that time a commonly-used approach to its resolution was based
on the introduction of the so-called ``biased'' (or asymmetric) vacuum.
As a result, under the appropriate choice of parameters, the regions of
``false'' (energetically unfavorable) vacuum should quickly disappear,
eliminating the corresponding domain walls~\cite{Gelmini89}.
Unfortunately, the concept of biased vacuum was not supported by
independent data in the physics of elementary particles.
On the other hand, the idea of nonlocal correlations, employed in
the present study, is supported by at least a few laboratory experiments.
Therefore, from our point of view, it looks more attractive.

It should be emphasized again that a number of arguments from the modern
observational cosmology impose severe constraints on the concentration of
domain walls. In particular, an appreciable number of the domain walls
would produce an unacceptable anisotropy of the cosmic microwave background
(CMB) radiation, change the overall rate of the cosmological expansion,
\textit{etc}. Besides, it is commonly believed now that the primordial
spectrum of density fluctuations, responsible for the formation of
the large-scale structure, is formed by the Gaussian quantum fluctuations
in the very early Universe amplified by the subsequent inflationary stage,
while contribution from the topological defects is quite insignificant.
All these facts imply that there should be an efficient mechanism for
the suppression of the domain walls, and the nonlocal quantum correlations
discussed in the present paper might be a reasonable option.

Yet another recent cosmological problem, recognized due to \textit{WMAP}
and confirmed by the \textit{Planck} satellite data, is the anomalous
behaviour of CMB fluctuations at large angular scales, approximately
over~$ 10^{\rm o} $~\cite{Wright13,Hofmann13}. It can be conjectured that
such anomalies are associated just with the nonlocal correlations in
the early Universe; but, of course, a much more elaborated analysis must
be performed to draw a reliable conclusion.

At last, we would like to mention a recent activity in the experiments with
ultracold gases for simulation of various dynamical phenomena in cosmology,
\textit{e.g.}, the so-called Sakharov oscillations~\cite{Schmiedmayer13}.
From our point of view, a more careful study of the nonlocal correlations
may become an important branch in this rapidly-growing research field.

\section*{Acknowledgments}

The initial stage of this work was substantially supported by the ESF COSLAB
(Cosmology in the Laboratory) Programme.

I am grateful to a large number of researchers with whom I discussed
the problem of nonlocal effects in cosmology since the early 2000s till now:
E.~Arimondo,
V.B.~Belyaev,
R.A.~Bertlmann,
Yu.M.~Bunkov,
A.M.~Che\-chelnitsky,
I.~Coleman,
J.~Dalibard,
V.B.~Efimov,
V.B.~Eltsov,
H.J.~Junes,
I.B.~Khriplovich,
T.W.B.~Kibble,
M.~Knyazev,
V.P.~Koshelets,
O.D.~Lavrentovich,
V.N.~Lukash,
A.~Maniv,
P.V.E.~McClintock,
L.B.~Okun,
G.R.~Pickett,
E.~Polturak,
A.I.~Rez,
R.J.~Rivers,
M.~Sakellariadou,
M.~Sasaki,
R.~Sch{\"u}tzhold,
V.B.~Semikoz,
M.~Shaposhnikov,
A.A.~Starobinsky,
A.V.~Toporensky,
W.G.~Unruh,
G.~Vitiello,
G.E.~Volovik,
C.~Wetterich, and
W.H.~Zurek.

The author declares that there is no conflict of interests
regarding the publication of this article.


\begin{thebibliography}{18}

\bibitem[{Einstein \textit{et~al.}}(1935)]{Einstein35}
A.~Einstein, B.~Podolsky, and N.~Rosen,
``Can quantum-mechanical description of physical reality
be considered complete?''
\emph{Physical Review}, vol.~47, no.~10, pp.~777--780, 1935.

\bibitem[{Linde}(1979)]{Linde79}
A.D. Linde,
``Phase transitions in gauge theories and cosmology,''
\emph{Reports on Progress in Physics},
vol.~42, no.~3, pp.~389--437, 1979.

\bibitem[{Bogoliubov}(1966)]{Bogoliubov66}
N.N.~Bogoliubov,
``Field-theoretical methods in physics,''
\emph{Supplemento al Nuovo Cimento (Serie prima)},
vol.~4, no.~2, pp.~346--357, 1966.

\bibitem[{Zel'dovich \textit{et~al.}}(1974)]{Zeldovich74}
Ia.B.~Zeldovich, I.Yu.~Kobzarev, and L.B.~Okun,
``Cosmological consequences of a spontaneous breakdown of
a discrete symmetry,''
\emph{Soviet Physics---JETP}, vol.~40, no.~1, pp.~1--5, 1975
[Translated from:
\emph{Zhurnal Eksperimental'noi i Teoreticheskoi Fiziki},
vol.~67, p.~3--11, 1974].

\bibitem[{Gelmini \textit{et~al.}}(1989)]{Gelmini89}
G.B.~Gelmini, M.~Gleiser, and E.W.~Kolb,
``Cosmology of biased discrete symmetry breaking,''
\emph{Physical Review D}, vol.~39, no.~6, pp.~1558--1566, 1989.

\bibitem[{Kibble}(1976)]{Kibble76}
T.W.B.~Kibble,
``Topology of cosmic domains and strings,''
\emph{Journal of Physics A: Mathematical and General},
vol.~9, no.~8, pp.~1387--1398, 1976.

\bibitem[{Zurek(1985)}]{Zurek85}
W.H.~Zurek,
``Cosmological experiments in superfluid helium?''
\emph{Nature}, vol.~317, no.~6037, pp.~505--508, 1985.

\bibitem[{Klapdor-Kleingrothaus and Zuber}(1997)]{Klapdor97}
H.V.~Klapdor-Kleingrothaus and K.~Zuber,
\emph{Particle Astrophysics},
Institute of Physics Publishing, Bristol, 1997.

\bibitem[{Carmi \textit{et~al.}}(2000)]{Carmi00}
R.~Carmi, E.~Polturak, and G.~Koren,
``Observation of spontaneous flux generation in
a multi-Josephson-junction loop,''
\emph{Physical Review Letters}, vol.~84, no.~21,
pp.~4966--4969, 2000.

\bibitem[{Hadzibabic \textit{et~al.}}(2004)]{Hadzibabic04}
Z.~Hadzibabic, S.~Stock, B.~Battelier, V.~Bretin, and J.~Dalibard,
``Interference of an array of independent Bose--Einstein condensates,''
\emph{Physical Review Letters}, vol.~93, no.~18,
Article ID~180403 (4~pp.), 2004.

\bibitem[{Hadzibabic \textit{et~al.}}(2006)]{Hadzibabic06}
Z.~Hadzibabic, P.~Kruger, M.~Cheneau, B.~Battelier, and J.~Dalibard,
``Berezinskii--Kosterlitz--Thouless crossover in a trapped
atomic gas,''
\emph{Nature}, vol.~441, no.~7097, pp.~1118--1121, 2006.

\bibitem[{Misner}(1969)]{Misner69}
C.W.~Misner,
``Mixmaster Universe,''
\emph{Physical Review Letters}, vol.~22, no.~20, pp.~1071--1074, 1969.

\bibitem[{Linde}(1984)]{Linde84}
A.D.~Linde,
``The inflationary Universe,''
\emph{Reports on Progress in Physics}, vol.~47, no.~8, pp.~925--986, 1984.

\bibitem[{Dumin}(2000)]{Dumin00}
Yu.V.~Dumin,
``On a probable role of EPR (Einstein--Podolsky--Rosen) correlations
in breaking the symmetry of Higgs fields in
cosmological phase transitions,''
\emph{Hot Points in Astrophysics:
Proceedings of the International Workshop}, pp.~114--120,
Joint Institute for Nuclear Research, Dubna, 2000.

\bibitem[{Isihara}(1971)]{Isihara71}
A.~Isihara,
\emph{Statistical Physics},
Academic Press, New York, 1971.

\bibitem[{Dumin}(2003)]{Dumin03}
Yu.V.~Dumin,
``On the influence of Einstein--Podolsky--Rosen effect on
the domain wall formation during the cosmological phase transitions,''
\emph{Frontiers of Particle Physics: Proceedings of the Tenth
Lomonosov Conference on Elementary Particle Physics},
pp.~289--294, World Scientific Publishing Co., Singapore, 2003.

\bibitem[{Dumin}(2009)]{Dumin09}
Yu.V.~Dumin,
``Ultracold gases and multi-Josephson junctions as simulators of
out-of-equilibrium phase transformations in superfluids and
superconductors,''
\emph{New Journal of Physics}, vol.~11, no.~10,
Article ID~103032 (12~pp.), 2009.

\bibitem[{Rumer and Ryvkin}(1980)]{Rumer80}
Yu.B.~Rumer and M.Sh.~Ryvkin,
\emph{Thermodynamics, Statistical Physics, and Kinetics},
Mir, Moscow, 1980.

\bibitem[{Wright}(2013)]{Wright13}
A.~Wright,
``Across the Universe,''
\emph{Nature Physics}, vol.~9, no.~5, p.~264, 2013.

\bibitem[{Hofmann}(2013)]{Hofmann13}
R.~Hofmann,
``The fate of statistical isotropy,''
\emph{Nature Physics}, vol.~9, no.~11, pp.~686--689, 2013.

\bibitem[{Schmiedmayer and Berges}(2013)]{Schmiedmayer13}
J.~Schmiedmayer and J.~Berges,
``Cold atom cosmology,''
\emph{Science}, vol.~341, no.~6151, pp.~1188--1189, 2013.

\end{thebibliography}
\end{document}